\documentclass[11pt]{emulateapj}
\usepackage{epsfig,apjfonts}
\def\Meszaros{M\'esz\'aros~}

\begin{document}

\title{Nonthermal $\gamma$/X-ray  flashes  from shock breakout  in gamma-ray bursts/supernovae }

\author{Xiang-Yu  Wang\altaffilmark{1,2},  Zhuo Li\altaffilmark{3},
Eli Waxman\altaffilmark{3} and Peter
M\'esz\'aros\altaffilmark{1,4} } \altaffiltext{1}{Department of
Astronomy and Astrophysics, Pennsylvania State University,
University Park, PA 16802, USA} \altaffiltext{2}{Department of
Astronomy, Nanjing University, Nanjing 210093, China}
 \altaffiltext{3}{Physics Faculty, Weizmann Institute of Science, Rehovot
76100, Israel } \altaffiltext{4}{Department of Physics,
Pennsylvania State University, University Park, PA 16802, USA}
\begin{abstract}
Thermal X-ray emission which is simultaneous with the prompt
gamma-rays has been detected for the first time from a supernova
connected with a gamma-ray burst (GRB), namely GRB060218/SN2006aj.
It has been interpreted as arising from the  breakout of a mildly
relativistic, radiation-dominated shock from a dense stellar wind
surrounding the progenitor star. There is also evidence for the
presence of a mildly relativistic ejecta in GRB980425/SN1998bw,
based on its X-ray and radio afterglow. Here we study the process
of repeated bulk Compton scatterings of shock breakout thermal
photons by the mildly relativistic ejecta. During the shock
breakout process, a fraction of the thermal photons would be
repeatedly scattered between the pre-shock material and the
shocked material as well as the mildly relativistic ejecta  and,
as a result,  the thermal photons get boosted to increasingly
higher energies. This bulk motion Comptonization mechanism will
produce nonthermal gamma-ray and X-ray flashes, which could
account for the prompt gamma-ray burst emission in low-luminosity
supernova-connected GRBs, such as GRB060218.  A Monte Carlo code
has been developed to simulate this repeated scattering process,
which confirms  that a significant fraction of the thermal photons
get ``accelerated" to form a nonthermal component, with a dominant
luminosity. This interpretation for the prompt nonthermal emission
of GRB060218 may imply that either the usual internal shock
emission from highly relativistic jets in these low-luminosity
GRBs is weak, or alternatively, that there are no highly
relativistic jets in this peculiar class of bursts.

\end{abstract}

\keywords{gamma rays: bursts --- supernovae: general --- shock waves}


\section{Introduction}
Supernova shock breakout has been predicted for a few decades
(e.g. Colgate 1974; Klein \& Chevalier 1978; Ensman \& Burrows
1992; Blinnikov et al. 1998, Matzner \& McKee 1999). In core
collapse supernovae, a shock wave is generated which propagates
through the progenitor star and ejects the envelope. As the shock
propagates through the envelope, it is mediated by radiation: The
post shock energy density is dominated by radiation, and the shock
transition is mediated by Compton scattering. As the shock
approaches the edge of the star, the optical depth of the plasma
lying ahead of the shock decreases. At the radius where the
optical depth drops to $\sim c/v_s$, where $v_s$ is the shock
velocity, Compton scattering can no longer sustain the shock, and
the shock undergoes a transition to a viscous shock (or a
collisionless shock, Waxman \& Loeb 2001). This transition is
accompanied by a bright ultraviolet/X-ray burst, as photons escape
far ahead of the shock due to the low optical depth. The term
"shock breakout" is commonly used to refer to the emergence of the
shock from the edge of the star. However, if the star is
surrounded by an optically thick wind, the radiation mediated
shock would continue to propagate into the wind, up to the point
where the wind optical depth drops below $\sim c/v_s$. In this
case, shock breakout, i.e. the transition from a radiation
mediated shock to a viscous (or collisionless) one accompanied by
a bright ultraviolet/X-ray burst, occurs as the shock propagates
through the wind, at a radius which may be significantly larger
than the star's radius. We use here the term "shock breakout" to
denote this transition in general, regardless of whether it occurs
as the shock reaches the edge of the star or further out within an
optically thick wind.

Shock breakout has previously never been directly detected from
any supernovae due to its transient nature and its very early
(minutes to hours) occurrence, in the absence of a suitably prompt
trigger alert. Recently, thanks to its sensitive gamma-ray trigger
and rapid slewing capability,  {\it Swift} has detected early
thermal X-rays emission from a supernova associated with a GRB,
namely GRB060218/SN2006aj (e.g. Campana et al. 2006; Pian et al.
2006; Mazzali et al. 2006; Modjaz et al. 2006; Sollerman et al.
2006; Mirabel et al. 2006;  Cobb et al. 2006; Soderberg et al.
2006; Liang et al. 2006a), which has been interpreted as arising
from the breakout of a radiation-dominated shock (Campana et al.
2006; Waxman et al. 2007).

The large energy carried by thermal X-rays, $\sim10^{49}$~erg, and
the long duration ($\sim 1000$~s) of the thermal X-ray emission
implies that shock breakout must occur at a radius of
$\sim10^{13}$~cm. The lack of Hydrogen lines implies, on the other
hand, a compact progenitor, a WR star of radius $\sim10^{11}$~cm.
This suggests that the progenitor was surrounded by an optically
thick wind, and that shock breakout occurred at a radius
$\sim10^{13}$~cm, where the optical depth of the wind becomes
small enough. The inferred mass loss rate,
$\sim3\times10^{-4}M_\odot/{\rm yr}$  (Campana et al. 2006) is
consistent with that typically expected for WR stars (e.g. Felli
\& Panagia 1982; Nugis \& Lamers 2000, 2002)\footnote{The
arguments based on radio observations against the presence of an
optically thick wind (Fan et al. 2006), are not conclusive, since
radio observations do not allow one to determine the explosion
parameters. In particular, the estimates for the kinetic energy
range between $\sim10^{48}$~erg and $\sim10^{50}$~erg, and the
ambient medium density estimates range between $\sim10^0{\rm
cm}^{-3}$ and $\sim10^2{\rm cm}^{-3}$ (Soderberg et al. 2006; Fan
et al 2006). }. Another possibility is that the shock breaks out
from an optically thick shell pre-ejected from the progenitor (the
required shell mass is only $10^{-7} M_\sun$, see Campana et al.
2006).

The simple result, that a wind photospheric radius
$R_{ph}\sim10^{13}$~cm is required to account for the thermal
emission as due to shock break out from a wind, is supported by
the calculations of Li (2007), who finds that the radius at which
the wind optical depth equals 20 is required to be
$R_*\sim100R_\odot$. Since the wind models investigated in Li
(2007) have $1.5\lesssim R_{ph}/R_*<3$ (compare his figures 12 and
3), his conclusion implies that $R_{ph}\sim10^{13}$~cm is
required, consistent with our conclusion. Note, however, that the
claim that $R_*\sim100R_\odot$ is required is not robust, since
the ratio $R_{ph}/R_*$ depends on the adopted wind velocity
profile, and is likely to be $>3$ for WR stars. { Note also that
when the mass loss is well above $10^{-4} M_\odot {\rm yr^{-1} }$,
the wind becomes very optically thick and the apparent stellar
radius grows, so that the star no longer looks like a WR star
(Nino Panagia, personal communication)}. It should also be pointed
out that the analysis of the required wind properties in Li (2007)
is far too restricted. A model is adopted by Li (2007) where the
shock is produced by a spherical explosion with fixed energy
($2\times10^{51}$~erg) and fixed ejecta mass ($2M_\odot$), and its
acceleration is due to propagation through the model wind profile.
Beyond the fact that dependence on the explosion parameters is not
explored, it is not at all clear that this is the correct
description of shock dynamics.  In fact, it is known that such
shock acceleration models (Matzner \& McKee 1999; Tan et al. 2001)
can not account for the energy deposited in the mildly
relativistic ejecta, $v/c=0.8$, observed in GRB 980425/SN1998bw
(Kulkarni et al. 1998, Waxman \& Loeb 1999, Chevalier \& Li 1999,
Waxman 2004). It could be, e.g., that this mildly-relativistic
shock is driven instead by a choked jet propagating through the
progenitor.

From the thermal energy density of the shock, Campana et al.
(2006) have inferred that the shell driving the
radiation-dominated shock in GRB060218/SN2006aj must be mildly
relativistic, with a velocity $\Gamma\beta\sim1-2$, where $\beta$
is the velocity in units of the speed of light $c$ and
$\Gamma=(1-\beta^2)^{-1/2}$ is the Lorentz factor. This shock
could be driven by the outermost parts of the envelope that get
accelerated to a mildly relativistic velocity when the supernova
shock accelerates in the density gradient of the envelope of the
supernova progenitor (Colgate 1974; Matzner \& McKee 1999; Tan et
al. 2001), probably a Wolf-Rayet (WR)  star. The more compact
configuration of a WR star, compared with a blue or red
supergiant, and the large explosion energy characterizing a
hypernovae may help to accelerate a considerable amount of matter
to a mildly relativistic, and probably anisotropic velocity
distribution (see however the comment above on SN1998bw).

Due to the connection with a GRB, there are also
other possibilities for the origin of this radiation-dominated
shock. In the collapsar model for GRBs, the shock could be driven
by the cocoon of a relativistic jet (M\'esz\'aros \& Rees 2001;
Ramirez-Ruiz et al. 2002; Zhang et al. 2003) or the slower
($\Gamma\beta\sim1-2$), high-latitude outer parts of a
relativistic jet. The emergence of the shock in this case would be
highly non-spherical. As pointed out in Camapana et al. (2006),
observations require a non-spherical shock breakout. This does not
necessarily imply a shock driven by a cocoon/jet, since the wind itself
may be highly non-spherical\footnote{We note in this context that the
claim in Ghisellini et al. (2006b), that the optical emission and the thermal
X-ray emission are not due to the same thermal component is not necessarily
valid, since Ghisellini et al. (2006b) assume spherical symmetry and neglect
in addition light travel time effects.}.
Observations of high optical polarization in SNe at early times of 3-5 days
(e.g. Gorosabel et al. 2006) are also consistent with non-spherical shocks.

The  relatively flat light curve of the X-ray afterglow of the
nearest GRB, GRB980425/SN1998bw, up to $\sim 100$ days after the
burst, has been argued to be caused by the coasting phase of a
mildly relativistic shell with energy of a few times $10^{49}\rm
erg$ (Waxman 2004). Similarly, the X-ray afterglow of another
nearby GRB, GRB031203/SN2003lw also had an early flat light curve
(Watson et al. 2004). Thus, it is possible that a mildly
relativistic ejecta is ubiquitous in supernovae-associated GRBs.
In this paper, we suggest that the nonthermal X-ray  and gamma-ray
emission in such objects  can arise from repeated bulk-Compton
scatterings (i.e. bulk motion Comptonization) of thermal {shock
breakout X-rays photons.}

Blandford \& Payne (1981) first noted the importance of bulk
motion acceleration of photons in a radiation-dominated shock.
They found that photons are preferentially upscattered by the bulk
motion rather than by the thermal motions of the electrons, and a
power-law spectrum extending to high energies forms, when the
electron thermal velocity is less than the shock velocity $v_s$.
Repeated scatterings using the energy of the bulk motions of two
approaching relativistic shells in the context of GRB internal
shocks was studied by Gruzinov \& \Meszaros (2000). They found
that the seed synchrotron photons can be boosted to much higher
energies, which is confirmed by their Monte Carlo simulations.
This process is equivalent to the Fermi acceleration mechanism of
particles (Blandford \&  Eichler 1987)  or  photon scattering off
Alvf\'{e}n waves (Thompson 1994), but here the mechanism, instead,
uses the relative bulk motion and accelerates photons. { The bulk
motion  Comptonization mechanism has also been invoked to
interpret the  high-energy spectra of  accreting black holes (e.g.
Shrader \& Titarchuk 1998; Titarchuk \& Shrader 2005).}

\section{The bulk motion Comptonization model}
Consider a mildly relativistic ejecta driving a
radiation-dominated shock into the stellar envelope of the GRB/SN
progenitor or into an optically thick wind (or pre-ejected shell)
surrounding it. Once the optical depth  of the material in front
of the shock drops below $c/v_s$ (where $v_s$ is the shock
velocity), the photons escape and produce a breakout flash. Since
the Thompson scattering optical depth is non-negligible in front
of the shock while the shock is breaking out, { some fraction of
the thermal photons will be scattered back. The back-scattered
photons will be scattered forward by the expanding ejecta or
shocked plasma, boosting up their energy. The backward-forward
scattering cycle may repeat itself
 many times for some fraction of the photons, boosting their energy by a large factor}.

For GRB060218/SN2006aj, the shock breakout may occur in the wind
or a pre-ejected shell, which serves as the scattering medium
target. The radius where  a mildly relativistic
radiation-dominated shock breaks out from a dense wind is
\begin{equation}
R_{\rm br}\simeq2\times10^{12}{\rm cm}
\left(\frac{\dot{M}}{10^{-4}M_{\sun}{\rm
yr}^{-1}}\right)\left(\frac{v_w}{10^8 {\rm cm
s^{-1}}}\right)^{-1},
\end{equation}
where $\dot{M}$ is the mass loss rate and $v_w$ is the wind
velocity. The optical depth of the ejecta itself at the shock
breakout radius  is
\begin{equation}
\tau_{ej}=\frac{\sigma_{\rm T}E_k}{4\pi(\Gamma-1) m_p c^2 R^2}=35
E_{k,50} \left(\frac{R_{\rm br}}{10^{13}\rm cm}\right)^{-2}
(\Gamma-1)^{-1},
\end{equation}
{ where $E_{k}=10^{50}E_{k,50}$~erg is the (isotropic equivalent)
kinetic energy of the mildly relativistic shell. For
$E_{k,50}\sim1$, $\tau_{ej}$ is much larger than unity, and the
ejecta can be regarded as a mirror which effectively reflects all
the photons incident upon it.}

{The post-shock electron temperature is determined by the
balance between heating by the ion-electron collisions and Compton
cooling by the thermal shock breakout photons. For a shock
velocity $v_s$, the ion temperature is $kT_i=(3/16)\mu v_s^2\simeq
9\times10^7 (v_s/c)^2 {\rm eV}$, where $\mu=m_p/2$ is the mean
atomic mass per particle for a plasma in which the thermal energy
is shared equally between protons and electrons (Waxman \& Loeb
2001). The ion-electron collision rate is
$\nu_{ie}=2\times10^{-2}(\rho/10^{-11}{\rm g
cm^{-3}})(kT_i/10^8{\rm eV})^{-3/2} \,{\rm s^{-1}}$ for $T_e<m_e
T_i/m_p$ (e.g. Krall \& Trivelpiece 1973), while the Compton
scattering cooling rate is $\nu_{\rm Comp}=(8\sigma_T/3m_e c)a
T_{th}^4=5000(kT_{th}/0.15{\rm KeV})^4 {\rm s^{-1}}$, where
$\rho\simeq4\rho_w=2\times10^{-11}({\dot{M}}/{10^{-4}M_{\sun}{\rm
yr}^{-1}})({v_w}/{10^8 {\rm cm s^{-1}}})^{-1} R_{{\rm br},
12}^{-2} {\rm g \,cm^{-3}}$ is the mass density of the shocked
wind, $\rho_w$ is the density of the pre-shocked wind and $T_{th}$
is the temperature of the thermal photons. The electron
temperature is determined by the balance between heating and
cooling $\nu_{ie}T_i=\nu_{\rm Comp}T_e$, which gives }
\begin{equation}
kT_e\sim0.3 \left(\frac{v_s}{c}\right)
\left(\frac{\rho}{10^{-11}\rm g cm^{-3}}\right)\left
(\frac{k_BT_{th}}{0.15{\rm KeV}}\right)^{-4} {\rm KeV}~.
\end{equation}
Thus, the electrons can be regarded as being essentially cold, in
comparison with the mildly relativistic bulk motion.

Below we first present in \S~2.1 { a qualitative description} of
the bulk comptonization emission, by analogy to the Comptonization
mechanism in thermal electron plasma. { This description is based
on results obtained assuming} a fixed optical depth $\tau$ for the
scattering medium in front of the shock. In reality, the optical
depth is decreasing with time due to shock expansion, { and may
therefore change considerably between repeated photon scattering.
In fact, the time scale for a single scattering cycle is typically
comparable to the shock expansion time, and only a small fraction
of the photons are scattered on a shorter time scale and gain
considerable energy. In \S~2.2 we present a discussion based on a
one-dimension Monte Carlo simulation of this process, which takes
into account the time dependence of the optical depth. {The
purpose of this simulation is to give a qualitative analysis of
the shock breakout bulk Comptonization mechanism, and to explore
whether the non-thermal component can be indeed dominant. A
detailed, full three-dimension Monte-Carlo simulation would
require a full understanding of the shock structure and the
anisotropic geometry (as discussed in \S~2.2 in more detail),
which is, however, largely unknown at present.} In
\S~2.3 we discuss the implications to GRB060218/SN2006aj.}

\subsection{A qualitative description}
{ Ignoring the time dependence of $\tau$, the optical depth ahead
of the shock, the physical situation is  similar to that of
Comptonization by a thermal electron plasma. In the present case,
the electrons are cold and their momentum is dominated by the bulk
motion.} Multiple scattering in a thermal electron plasma can lead
to a power-law spectrum of the scattered photons (Pozdnyakov et
al. 1976; 1983), which has found application in many astrophysical
contexts. {   Dermer, Liang \& Canfield (1991) further extended
this to a mildly relativistic thermal plasma.} Assuming each
scattering amplifies the photon energy by a factor $A$, the energy
of a photon escaping after $k$ scatterings is
$\varepsilon_k=\varepsilon_i A^k$, where $\varepsilon_i$ and
$\varepsilon_k$ are the initial and final photon energies
respectively. { For fixed $\tau$,} a photon scattered by the
ejecta has a probability $1-e^{-\tau}$ to be scattered back
towards the ejecta, and a probability $e^{-\tau}$ to escape. The
probability for a photon to undergo $k$ scatterings before
escaping is $(1-e^{-\tau})^k$, and since the photon energy is
multiplied by $A$ per scattering, the escaping photon intensity
would have a power-law shape

\begin{equation}
F(\varepsilon_k)\sim F(\varepsilon_i)(1-e^{-\tau})^k\sim
F(\varepsilon_i)(\varepsilon_k/\varepsilon_i)^{-\alpha}
\end{equation}
with
\begin{equation}
\alpha=-{\rm ln}(1-e^{-\tau})/{\rm ln}A.
\end{equation}
The  photon energy amplification factor $A$ is determined by the
kinetic energy of the electrons. For trans-relativistic electrons
{ and isotropic photon distributions,
$A\sim\Gamma^2(1+\beta^2/3)$}. The power-law spectrum extends to a
cutoff energy, { which is the smaller of the electron kinetic
energy, $\sim(\Gamma-1)m_e c^2$, and its rest mass $m_ec^2$ (due
to the Klein-Nishina effect). }

An important difference between the usual thermal electron (or
bulk) Comptonization case and the current case is that in the
present case the scattering optical depth decreases with time as
the mildly relativistic ejecta moves outward. Initially, when the
optical depth $\tau\ga1$, the slope of $\nu F_\nu$ is positive and
most radiation is emitted at high energies. As $\tau$ decreases,
the spectrum becomes softer and softer, and at late times the
spectrum is composed of a thermal peak plus a weak high-energy
power-law tail. In general, we expect a noticeable spectral
softening of the nonthermal emission with time. This spectral
softening is expected to be accompanied by a decrease in the
Compton luminosity. At early time, when the effective Compton
parameter $Y=A(1-e^{-\tau})>1$, the Compton luminosity may exceed
the thermal luminosity (it is limited by the kinetic energy of the
ejecta $E_k$). We expect the Compton luminosity to decrease with
time, as $Y$ decreases.

The gamma-ray light curves produced in this model generally have a
simple profile without multi-peak structure. The characteristic
variability timescale $\delta t$ of the burst is determined by the
the radius $R$ where the optical depth of the material ahead of
the shock drops to $\sim1$, i.e. $\delta t\sim R(\tau=1)/c$ (If
the stellar wind surrounding the progenitor were optically thin
everywhere, the shock would break out from the SN progenitor
stellar envelope and the variability time would be about
$R_{\star}/c$, where $R_{\star}$ is the stellar radius.)

\subsection{Monte Carlo simulation of photon ``acceleration''}
As the shock propagates outward, the optical depth $\tau$ ahead of
the shock decreases with time. In order to understand the photon
``acceleration'' mechanism in this time-dependent case, we carried
out a Monte Carlo simulation of repeated Compton scattering during
shock breakout from a dense stellar wind. {We approximate the
hydrodynamics of the problem as follows. We consider the mildly
relativistic ejected shell to act as a piston with a
time-independent bulk Lorentz factor $\Gamma$ and an infinite
optical depth $\tau_{ej}$. Since the stellar wind swept up by the
ejecta is not sufficient to decelerate it, the constant velocity
of the interface is justified. This``piston" drives a shock into
the surrounding medium, where the density profile is assumed to
follow $n \propto R^{-2}$. We assume the shock width to be
infinitesimal. There are three distinct regions in this picture:
The moving piston, the shocked medium and the pre-shocked medium.
The shocked medium is considered to form a homogeneous shell, and
the electrons are regarded as cold, as follows from Eq.~(3), with
a bulk velocity same as that of the ejecta. The velocity of the
shock front can then be obtained consistently from the shock jump
condition.   We further simplify the problem by considering a
one-dimensional situation, where the motion of the photons is
confined to one dimension, forward or backward relative to the
shock expansion direction. The photons are injected at the shock
front and would then be repeatedly scattered among these three
components until they escape out to a sufficiently further region.
Our simulation takes into account the proper possibility of photon
scattering by each of the three components (with the appropriate
photon energy gain or loss). The photon scattering probability by
the unshocked wind and shocked wind is determined by the optical
depth of each component, while the piston is regarded as a mirror
since it has a larger optical depth much larger than unity.}

{We note here that  this level of  simplification (i.e. the
one-dimensional treatment) is  appropriate for the uncertainties
inherent in such a scenario. Aside from the geometrical
uncertainties associated with the non-spherical scenario,
discussed further below, even in a spherical approximation a
complete calculation of the light curve and spectrum of escaping
photons would require a detailed calculation of the shock
structure, including the back-reaction of ``accelerated" photons
on the shock-the angular and energy distribution of electrons and
photons across the shock transition, which would determine the
distribution of ``photon injection times" and the distribution of
electron Lorentz factors across the shock. In order to do this, it
would not be sufficient to carry out a 3-D Monte Carlo calculation
of the Comptonization, where the photons are treated as test
particles. Rather, the structure of the shock and the radiation
field would need to be solved self-consistently, since the
radiation field affects the shock structure (electron energy
distribution etc). While such solutions exist for non relativistic
shocks (Weaver 1976), the structure of radiation-dominated shocks
is not known even for mildly relativistic shocks ($v/c>0.1$).
Constructing such a self-consistent solution would be beyond the
scope of the present manuscript.}
\begin{figure}[t]
\centering \epsfig{figure=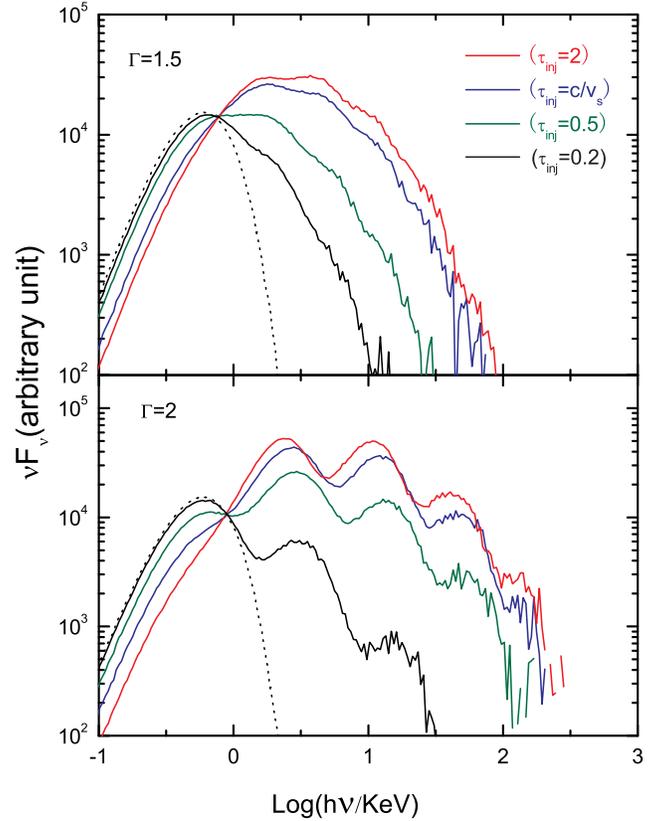,width=10cm} \caption{{ The
time-integrated energy distribution of the escaping photons.
$10^6$ photons, with a black body distribution at $k_B
T_{th}=0.15{\rm KeV}$ (black dotted line), are injected at four
different times, corresponding to optical depths of the
pre-shocked medium of $\tau_{\rm inj}=2$, $c/v_s$, 0.5 and 0.2.
The ejecta Lorentz factors are $\Gamma=1.5$ and $\Gamma=2$ for
upper panel and lower panel respectively. Note, that the ``humps''
seen in the spectra are an artifact of the one-dimensional
simulation, and are expected to be smoothed out in reality.}}
\end{figure}

\begin{figure}[t]
\centering \epsfig{figure=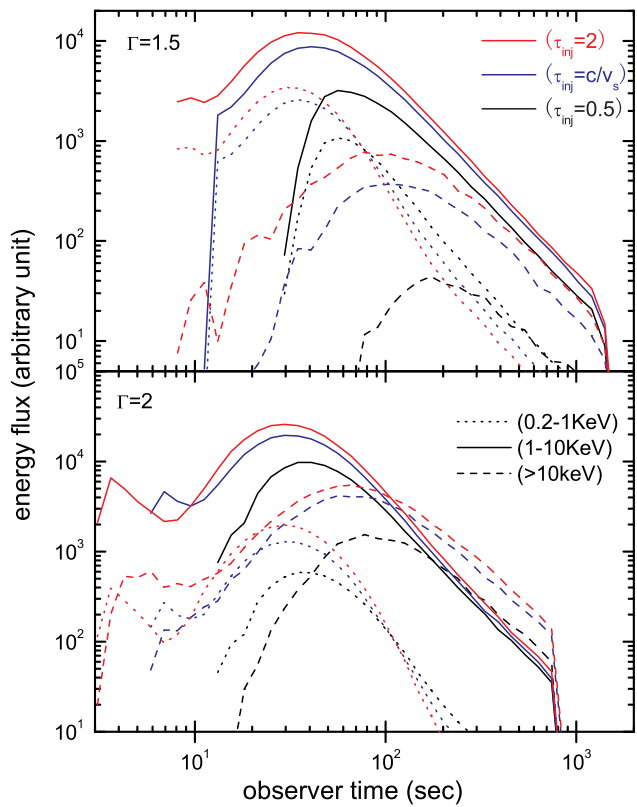,width=10cm} \caption{The temporal
evolution of the flux of "accelerated photons" {resulting from the
injection of photons at a single given time, corresponding to a
given value of $\tau_{\rm inj}$, for a simplified planar shock
breakout in one-dimension and two values of $\Gamma$}. The wind
parameters are: $\dot{M}=10^{-4} M_{\sun} {\rm yr}^{-1}$ and
$v_{w}=10^{8} {\rm cm s^{-1}}$. The injected photon temperature is
$k_B T_{th}=0.15{\rm KeV}$.  {Note that these curves are {\it not}
the light curve. The light curve in this model would be produced
by a convolution of the temporal profiles given in the figure with
the temporal dependence of the injected photon flux and energy,
its total duration being of the same order as that of the light
curve for the thermal photons, $\sim 10^3$ s, in the case of
GRB060218.}}
\end{figure}
We study the spectra of escaping photons, which result from the
``injection'' of photons at the shock front at various radii,
corresponding to various values of $\tau$ at the injection time,
denoted by $\tau_{\rm inj}$. The photons are treated as ``test
particles'', and their scattering history is followed as the shock
expands and $\tau$ decreases. The resulting time-integrated $\nu
F_\nu$ spectra are given in Fig.~1 for two values of $\Gamma$,
$\Gamma=2$ and $\Gamma=1.5$. The red, blue, olive and black curves
describe the spectrum of escaping photons resulting from the
injection of a thermal distribution of photons at $\tau_{\rm
inj}=2$, $c/v_s$, $0.5$ and $0.2$ respectively. The injected
photons are assumed to have a thermal spectrum with $T=0.15$~keV.
The "humps" seen in the simulated spectra correspond to different
orders of Compton scattering, with two nearby humps separated by
$\sim\Gamma^2$. These humps appear because we do not consider in
our one-dimensional simulation the angular distribution of
scattered photons. In reality, these humps are expected to be
smoothed out. Fig.~1 demonstrates that a significant fraction the
thermal photons that are injected at optical depth $\tau_{\rm
inj}\sim1$, where the photons are expected to escape the shock,
may be "accelerated" to high energy. The resulting non-thermal
component may carry a significant fraction of the shell energy,
and its luminosity may exceed that of the thermal component. As
expected, the spectum depends on $\Gamma$ and on $\tau_{\rm inj}$,
with harder spectra obtained for larger values of $\Gamma$ and
$\tau_{\rm inj}$. Note that the spectra are steeper (softer) at
high energies, due to the longer time required for acceleration to
higher energies, which implies a significant decrease in $\tau$
during the acceleration process. The peak in $\nu F_\nu$ is
typically at few keV, and the spectrum is essentially cutoff
beyond few hundred keV.

Fig.~2 shows the arrival time distribution of accelerated photons
in different energy channels, {for one single-time injection
in a one-dimensional planar shock breakout,  with particular
values for $\Gamma$ and $\tau_{\rm inj}$.} The higher energy
photons (e.g. $>10$ keV, dashed lines) are delayed compared to the
lower energy ones (e.g. 1-10 keV, solid lines of the same color).
For the wind parameters chosen, $\dot{M}=10^{-4} M_{\sun} {\rm
yr}^{-1}$ and $v_{w}=10^{8} {\rm cm s^{-1}}$, the delay is of
order tens of seconds (for photons injected at the same $\tau_{\rm
inj}$), and the decay time of the Compton flux is $\sim 100 ~{\rm
s}$. {Note that these curves are {\it not} the light curve.
The light curve in this model would be produced by the convolution
of the temporal profiles given in the figure with the temporal
dependence of injected photon flux and energy. Thermal photons
injected at different $\Gamma$, $\tau$ and different angles from
the line of sight, depending on the anisotropic geometry of the
radiation-dominated shock, will be continuously injected over a
time much longer than the individual pulses seen in Fig. 2. In the
following section, we apply these results to the case of
GRB060218/SN2006aj.}

\subsection{GRB060218/SN2006aj}

{The spectral analysis of GRB060218 shows that the spectrum
from X-ray to gamma-ray energies can be modelled as the sum of a
thermal component plus a  power-law with a high-energy cutoff
(Campana et al. 2006; Liang et al. 2006) and that the non-thermal
emission detected by Swift BAT and XRT is likely to be of the same
origin (Liang et al. 2006)}. This is consistent with the
theoretical expectations from the Comptonization model discussed
above. In particular, the location of the $\nu F_\nu$ peak at a
few keV is consistent with X-ray flash nature of GRB060218. As
pointed out by Dai et al (2006), the non-detection of an optical
emission of the synchrotron tail of the nonthermal X-ray emission
is an argument against the synchrotron internal shock origin for
the nonthermal emission. This lack of nonthermal optical emission
is naturally accounted for in our model, since repeated Compton
scattering can only boost thermal photons to higher energies.

Since the characteristic evolution time of GRB060218 is $\sim1000
~{\rm s}$, the $\sim 100 ~{\rm s}$ spread introduced by the
repeated scattering in our model is unimportant. The $\sim1000 ~
{\rm s}$ evolution timescale of the X-ray emission of GRB060218
has been argued to be dominated by light travel time effects,
possibly enhanced by the lateral dynamic time of a non-spherical
shock expansion (Campana et al. 2006; Waxman et al. 2007). {As
pointed out in Waxman et al. (2006), the mildly-relativistic
ejecta in GRB060218/SN2006aj is likely to be anisotropic, either
due to an anisotropic explosion, or due to being driven by a jet.
This, together with an anisotropic wind profile caused by rotation
(Meynet \& Maeder 2007) should lead to significant departures from
sphericity in the shock propagation. In an anisotropic shock, the
timescale is no longer the  spherical value $r/c\sim 200 ~{\rm
s}$, but is rather given by the sideways pattern expansion
timescale, which depends on the angular velocity profile of the
anisotropic shell (e.g. at larger angles the shock emerges later
due to a decreasing velocity profile or due to an increasing wind
density away from the symmetry axis, etc.). In any case, the
duration of the non-thermal part should be comparable to that of
the thermal part, which is of the order of $\sim 1000$ s. } In
this context, the temporal evolution of GRB060218 would reflect
the fact that, at different times, we see radiation from different
regions of the shock breakout that are characterized by different
$\Gamma$ and $\tau$. {The light curve  would be produced by
the convolution of the temporal profiles given in the figure with
the temporal dependence of injected photon flux and energy}. We
point out that a complete calculation of the light curve would
require a knowledge of the intensity of the thermal photons at
different times (corresponding to different $\tau_{\rm inj}$) and
of their temperatures. In order to derive these quantities, a
detailed calculation of the shock structure would be required,
which is beyond the scope of the present paper.

\section{Discussion}

We have discussed a model for the early gamma--ray and X-ray
emission of low luminosity GRBs which are associated with
supernovae, based on bulk Comptonization of thermal photons by a
semi-relativistic ejecta. One expects the following basic
characteristics for the gamma-ray bursts produced through the
above mechanism:

1) Smooth light curve profiles. The production of the gamma-rays
is through the repeated bulk-Compton scattering of thermal photons
emitted by the radiation dominated shock. The light curve should
follow the time behavior of the shock breakout and should
therefore be smooth. The light curves would generally have a
simple profile without multi-peak structure. The three nearby
low-luminosity bursts, GRB060218, GRB980425 and GRB031203, all
have smooth light curves, consistent with this picture.  Given
that GRB980425 and GRB031203 are also believed to have produced
mildly relativistic ejecta, it is possible that the gamma-ray
emission is due to supernova shock breakout also in these two
bursts. { Based on the association between GRB980425 and SN1998bw,
Bloom et al. (1998) speculated that there may exist a subclass of
smooth light curve GRBs produced by SNe.} The duration of
GRB980425 and GRB031203 is only tens of seconds, which can be
interpreted as due to shock break out from the progenitor
envelope, implying that the wind surrounding the progenitor is
optically thin. The characteristic variability time of GRB980425
and GRB031203 is a few seconds, corresponding to
$R(\tau=1)\sim10^{11}\rm cm$, which is comparable to the stellar
radius of a Wolf-Rayet star. The speculation that the stellar wind
surrounding  GRB980425 is optically thin is consistent with the
wind mass-loss rate inferred from the X-ray and radio afterglow of
GRB980425/SN1998bw (Waxman 2004; Li \& Chevalier 1999). The longer
timescale of GRB 060218 can, on the other hand, be interpreted in
terms of the shock breaking out from an optically thick wind.

\begin{table*}
\caption{The spectrum of three nearby low-luminosity GRBs}
\begin{center}
\begin{tabular}{lllllllll}
\hline \hline
GRB/SN & z & $E_{\gamma, {\rm iso}}($\rm erg$) $ &  $\alpha$& $\varepsilon_c (\rm KeV)$ &  references$^*$\\
\hline GRB980425/SN1998bw & 0.0085&$(8.5\pm0.1)\times10^{47}$ &$0.45\pm0.22$ & $\sim 200$ &1;3 \\
GRB031203/SN2003lw & 0.105 & $(4\pm1)\times10^{49}$ &$0.63\pm0.06$ &$>190$ & 2;4\\
 GRB060218/SN2006aj & 0.0331& $(6.2\pm0.3)\times10^{49}$ &0.45 & $\sim 30$$^\S$& 1;3;5; 6\\
\hline
\end{tabular}
\end{center}

$^{*}$The redshift $z$, spectral index $\alpha$ ($F_\nu\propto
\nu^{-\alpha}$) and cutoff energy $\varepsilon_c$ of the three
nearby, sub-luminous SN-connected GRBs. References: (1) Ghisellini
et al. (2006); (2) Sazonov et al. 2004;
(3)Kaneko et al. 2006; (4) Sazonov et al. 2004; (5) Campana et al. 2006; (6) S. Campana, private communication. \\
$^\S$ For the time $<300$ seconds after the trigger.
\end{table*}

2)The spectrum is expected to be composed of a simple power-law,
with a high energy cutoff lower than few $\times100$~keV, and a
thermal X-ray component. The spectrum of the gamma-ray emission
can be modelled as a cutoff power law rather than the Band
function for the usual bursts. The spectrum is expected to evolve
from hard to soft since the spectral index becomes smaller and the
cutoff energy decreases as the optical depth decreases. The
spectra of three nearby SN-GRBs, indicated in Table 1, are
consistent with this cutoff power-law spectrum.

Cosmological GRBs with isotropic equivalent energes
$E\sim10^{51}-10^{54} {\rm erg}$ are generally believed to be
produced through internal shocks in relativistic jets with Lorentz
factor $\Gamma\ga 100$ (for a recent review, see \Meszaros 2006).
In the collapsar scenario, relativistic jets can break free from
the star  along the rotation axis of a collapsing stellar core,
provided that the central engine feeding time of the jets is
sufficient long. Light curve breaks in the afterglow emission
attributed to jet effects have been seen in many bursts. On the
other hand, three nearby GRBs, GRB980425/SN1998bw,
GRB031203/SN2003lw and GRB060218/SN2006aj, have isotropic energies
in the range $\sim10^{48}-10^{49}{\rm erg}$, much lower than
typical cosmological bursts. Due to their proximity, the inferred
intrinsic rate of these sub-energetic events is, however, much
higher (Soderberg et al. 2006; Liang et al. 2006b), raising
difficulties in interpreting these events as typical GRBs observed
off-axis (Cobb et al. 2006). Both the properties of the prompt
gamma-ray emission and of the afterglows are different from those
of typical cosmological bursts. Up to now there is no
straightforward evidence for the presence of highly relativistic
jets in these sub-energetic bursts, but instead, there does exist
evidence for mildly relativistic ejecta in these bursts. Thus, it
could be that the relativistic jets in these events are much
weaker, giving most of their energy to the outer slow material
when they are burrowing their way though the star, which leads to
a much wider anisotropic pattern of outflow. This could be the
case even if the jets are chocked before emerging outside the
stellar envelope, due e.g. to an insufficient  central engine
feeding time. Whether there is a relativistic jet or not can be
probed through sub-GeV to GeV observations during the shock
breakout (Wang \& \Meszaros 2006)

Mildly relativistic ejecta may also exist in the usual
high-luminosity long GRBs, since  shock acceleration is expected
to accompany the supernova. This might explain the low-velocity
component, other than the relativistic jet, found by Berger et al.
(2003) and Kartik et al. (2003) in another nearby
supernova-connected GRB, namely GRB030329/SN2003dh, from the radio
and optical afterglow more than 1.5 days after the explosion. In
this burst, the gamma-ray emission from the mildly relativistic
ejecta is lower than that of the relativistic jet, since the
latter has a much higher isotropic energy. As suggested by Berger
et al (2003), the total energy in the low velocity component and
the relativistic jet may be a roughly constant quantity, the only
variable being the ratio of these two components. In this vein, we
would predict that at high redshifts there could be many GRB/SN
with stronger lower velocity and weaker relativistic jet
components, which would be harder to detect, even though their
rates of occurrence might exceed that of the more easily
detectable bursts with weak low-velocity and stronger relativistic
jets.

{\acknowledgments We would like to thank S. Campana, Z. G. Dai, B.
Zhang, E. W. Liang and Nino Panagia for useful discussions. This
work is partially supported by NASA NAG5-13286, NSF AST 0307376,
the National Natural Science Foundation of China under grants
10403002, 10221001 and 10473010, and the Foundation for the
Authors of National Excellent Doctoral Dissertations of China (for
XYW). ZL \& EW are partially supported by ISF and AEC grants.}

\end{document}